\documentclass[aps,prb,twocolumn,notitlepage,floatfix,superscriptaddress]{revtex4-1}

\usepackage{lmodern}
\usepackage{graphicx}
\usepackage{bm}
\usepackage{amsmath}
\usepackage{amssymb}
\usepackage{color}
\usepackage{pdfsync}
\usepackage[percent]{overpic}
\usepackage[breaklinks=true,colorlinks,citecolor=blue,linkcolor=blue,urlcolor=blue]{hyperref}
\usepackage[normalem]{ulem}

\begin{document}

\title{Electrical plasmon detection in graphene waveguides}

\author{Iacopo Torre}
\affiliation{NEST, Istituto Nanoscienze-CNR and Scuola Normale Superiore, I-56126 Pisa,~Italy}

\author{Andrea Tomadin}
\affiliation{NEST, Istituto Nanoscienze-CNR and Scuola Normale Superiore, I-56126 Pisa,~Italy}

\author{Roman Krahne}
\affiliation{Istituto Italiano di Tecnologia, Graphene Labs and Nanochemistry Department, Via Morego 30, I-16163 Genova,~Italy}

\author{Vittorio Pellegrini}
\affiliation{Istituto Italiano di Tecnologia, Graphene Labs, Via Morego 30, I-16163 Genova,~Italy}
\affiliation{NEST, Istituto Nanoscienze-CNR and Scuola Normale Superiore, I-56126 Pisa,~Italy}

\author{Marco Polini}
\email{m.polini@sns.it}
\affiliation{NEST, Istituto Nanoscienze-CNR and Scuola Normale Superiore, I-56126 Pisa,~Italy}
\affiliation{Istituto Italiano di Tecnologia, Graphene Labs, Via Morego 30, I-16163 Genova,~Italy}

\begin{abstract}
We present a simple device architecture that allows all-electrical detection of plasmons in a graphene waveguide.
The key principle of our electrical plasmon detection scheme is the non-linear nature of the hydrodynamic equations of motion that describe transport in graphene at room temperature and in a wide range of carrier densities.
These non-linearities yield a dc voltage in response to the oscillating field of a propagating plasmon.
For illustrative purposes, we calculate the dc voltage arising from the propagation of the lowest-energy modes in a fully analytical fashion.
Our device architecture for all-electrical plasmon detection paves the way for the integration of graphene plasmonic waveguides in electronic circuits.
\end{abstract}

\maketitle
{\it Introduction.}---The two-dimensional (2D) electron liquid in a doped graphene sheet~\cite{kotov_rmp_2012} supports plasmons with energies from the far-infrared to the visible, depending on carrier concentration~\cite{Diracplasmons}.
Although they share similarities with plasmons in ordinary parabolic-band 2D electron liquids~\cite{Giuliani_and_Vignale}, plasmons in graphene are profoundly different.
From a fundamental point of view, their dispersion relation is sensitive to many-body effects even in the long-wavelength limit~\cite{galileaninvariance}.
More practically, plasmons in graphene are easily accessible to surface-science probes and opto-electrical manipulation since they are exposed and not buried in a quantum well.

Plasmons in graphene are also substantially different from those in noble metals.
Indeed, recent near-field optical spectroscopy experiments~\cite{fei_nanolett_2011,fei_nature_2012,chen_nature_2012,alonso_science_2014,woessner_arxiv_2014} have demonstrated that plasmons in graphene display gate tunability and ultra-strong field confinement.
Moreover, low damping rates can be achieved by employing graphene samples encapsulated in hexagonal boron nitride thin slabs~\cite{principi_prb_2013,principi_prbr_2013,principi_prb_2014,woessner_arxiv_2014}.
\begin{figure}
\includegraphics[width=0.80\linewidth]{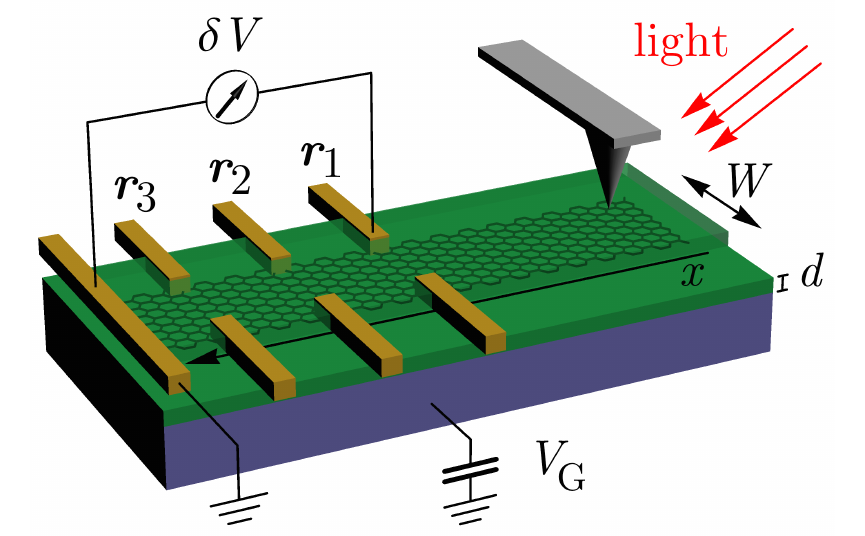}
\caption{(Color online) Schematics of our electrical plasmon detector.
A graphene strip of width  $W$ is encapsulated between two dielectrics (semi-transparent slab above and dark green slab underneath graphene).
A back gate (dark blue slab), separated by a distance $d$ from the graphene sheet and held at a voltage $V_{\rm G}$, is used to control the average carrier density ${\bar n}$ in  graphene.
At one end of the strip, a plasmon is launched by using e.g~a metallized atomic force microscope tip illuminated by light~\cite{fei_nature_2012,chen_nature_2012,woessner_arxiv_2014,fei_nanolett_2011,alonso_science_2014}.
Due to non-linearities in the hydrodynamic equations, a dc electrical potential difference $\delta V$ is measured between probe electrodes placed at positions ${\bm r}_1$, ${\bm r}_2$, and ${\bm r}_3$ and a reference electrode placed at the other end of the strip. The quantity $\delta V$ provides a direct measurement of the ac electric field of a propagating plasmon.\label{fig:setup}}
\end{figure}
For these reasons, graphene plasmonics has recently attracted a great deal of interest~\cite{reviews}.
Graphene plasmons may allow for new classes of devices for single-plasmon non-linearities~\cite{gullans_prl_2013}, extraordinarily strong light-matter interactions~\cite{koppens_nanolett_2011}, deep sub-wavelength metamaterials~\cite{ju_naturenano_2011,yan_naturenano_2012,brar_nanolett_2013,fang_nanolett_2014}, and photodetectors with enhanced sensitivity~\cite{koppens_naturenano_2014,THzdetectors}.

A key ingredient of a disruptive plasmonic platform is the ability to efficiently detect plasmons in all-electrical manners.
Some progress has been made in this direction in conventional noble-metal-based plasmonics.
Falk {\it et al.}~\cite{falk_naturephys_2009}, for example, were able to couple plasmons in Ag nanowires to nanowire Ge field-effect transistors.
Built-in electric fields in the latter are used to separate electrons and holes before recombination, thereby giving rise to a measurable source-drain current.
Similarly, Neutens {\it et al.}~\cite{neutens_naturephoton_2009} employed an integrated metal-semiconductor-metal detector in a metal-insulator-metal plasmon waveguide.

While graphene plasmons have been detected and studied in a multitude of ways~\cite{reviews}, including electron energy loss spectroscopy~\cite{EELS}, polarized Fourier transform infrared spectroscopy~\cite{ju_naturenano_2011}, and near-field optical spectroscopy~\cite{fei_nature_2012,chen_nature_2012,woessner_arxiv_2014,fei_nanolett_2011,alonso_science_2014}, a protocol for all-electrical detection of these modes is still lacking.

In this work we present a device architecture that allows all-electrical detection of plasmons in graphene waveguides.
In our scheme, all-electrical detection is not enabled by the integration of a detector in a graphene plasmon waveguide (GPW) but rather by the {\it intrinsic} non-linear terms in the hydrodynamic equations that describe transport in the 2D massless Dirac fermion (MDF) liquid~\cite{kotov_rmp_2012} hosted by graphene.
Non-linearities enable the emergence of a rectified (i.e.~dc) component $\delta V({\bm r})$ of the ac electric field of a propagating plasmon, which can be measured by a suitable geometry of ohmic contacts placed along the GPW, as shown in Fig.~\ref{fig:setup}.
We now present a calculation of the spatially-dependent electrical signal $\delta V({\bm r})$.

{\it Hydrodynamic theory.}---We consider a GPW with transverse (longitudinal) size $W$ ($L$ with $L\gg W$), which is embedded between two insulators with dielectric constants $\epsilon_1$ (above the GPW) and $\epsilon_2$ (below the GPW). Here, ``longitudinal'' and ``transverse'' refer to the plasmon propagation direction---${\hat {\bm x}}$ in Fig.~\ref{fig:setup}.

We would like to describe ac transport in a GPW by employing the theory of hydrodynamics~\cite{landau06}. We therefore need to assess whether experimentally relevant regions of parameter space exist in which this theory is applicable. First, at room temperature and for typical carrier densities (${\bar n} \simeq 10^{11}~{\rm cm}^{-2}$-$5 \times 10^{12}~{\rm cm}^{-2}$), the mean-free-path $\ell_{\rm ee} = v_{\rm F} \tau_{\rm ee}$ for electron-electron collisions in graphene is short~\cite{li_prb_2013,polini_arxiv_2014}, i.e.~$\ell_{\rm ee} \simeq 100$-$150~{\rm nm}$.
Here, $v_{\rm F} \simeq 10^6~{\rm m}/{\rm s}$ is the graphene Fermi velocity~\cite{castroneto_rmp_2009} and $\tau_{\rm ee} \simeq 100~{\rm fs} = 10^{-13}~{\rm s}$ is the electron-electron collision time~\cite{li_prb_2013,polini_arxiv_2014}. Second, for hydrodynamics to provide a correct description of the response of the system at finite frequencies, it must also be $\omega \tau_{\rm ee} \ll 1$, where $\omega$ is the external-excitation angular frequency. The value of $\tau_{\rm ee}$ given above constraints the maximum external-excitation frequency to be $f_{\rm max} \equiv 1/(2\pi \tau_{\rm ee}) \lesssim 3~{\rm THz}$. We therefore conclude that, for ${\bar n} \simeq 10^{11}~{\rm cm}^{-2}$-$5 \times 10^{12}~{\rm cm}^{-2}$, $\omega < 2\pi f_{\rm max}$, and $T =300~{\rm K}$, transport in GPWs with characteristic dimensions $L,W \gg \ell_{\rm ee}$ is accurately described by hydrodynamic equations of motion~\cite{landau06}.
Related continuum-model descriptions of plasmons in GPWs have been employed in Refs.~\onlinecite{wang_prb_2011,nikitin_prb_2011,christensen_acsnano_2012,wang_prb_2013}.

The set of hydrodynamic equations consists of i) the continuity equation,
\begin{equation}\label{eq:continuity}
\partial_t n({\bm r}, t)+\nabla \cdot \left[ n({\bm r},t) {\bm v}({\bm r},t)\right]=0~,
\end{equation}
and ii) the Navier-Stokes equation~\cite{landau06}
\begin{equation}\label{eq:NavierStokes}
m_{\rm c} n({\bm r}, t)D_t{\bm v}({\bm r}, t) = - e n({\bm r},t){\bm E}({\bm r}, t) + \eta \nabla^2{\bm v}({\bm r},t)~.
\end{equation}
In Eqs.~(\ref{eq:continuity})-(\ref{eq:NavierStokes}), $n({\bm r},t)$ is the carrier density and ${\bm v}({\bm r},t)$ is the drift velocity.
In Eq.~(\ref{eq:NavierStokes}), $m_{\rm c}=\hbar \sqrt{\pi \bar{n}}/v_{\rm F}$ is the graphene cyclotron mass~\cite{castroneto_rmp_2009}, with $\bar{n} = C V_{\rm G}/e$ the average electron density and $V_{\rm G}$ the back-gate voltage (see Fig.~\ref{fig:setup}), and $D_t \equiv \partial_t  + {\bm v}({\bm r}, t)\cdot \nabla$ is the convective derivative~\cite{landau06}.
The electric field ${\bm E}({\bm r}, t) = - \nabla\Phi({\bm r}, t)$ is the gradient of the electrostatic potential $\Phi({\bm r}, t)$ (we neglect retardation effects).
Finally, $\eta$ is the shear viscosity of the 2D electron liquid~\cite{Giuliani_and_Vignale,landau06}. For future purposes, we also introduce the kinematic viscosity~\cite{landau06}
\begin{equation}\label{eq:kinematic}
\nu \equiv \frac{\eta}{\bar{n}m_{\rm c}}~.
\end{equation}
It can be shown~\cite{principi_shearviscosity_2014}  that, in the hydrodynamic $\omega \tau_{\rm ee}\ll 1$ limit, 
$\nu \simeq v_{\rm F}^2 \tau_{\rm ee}/4$. With the values of $v_{\rm F}$ and $\tau_{\rm ee}$ given above, we find $\nu \simeq 250~{\rm cm}^2/{\rm s}$. In writing Eq.~(\ref{eq:NavierStokes}) we have neglected a term due to the bulk viscosity $\zeta$ since this quantity vanishes at long wavelengths~\cite{Giuliani_and_Vignale,landau06}. 

We highlight {\it two} non-linear terms in Eqs.~(\ref{eq:continuity})-(\ref{eq:NavierStokes}): a) the non-linear coupling between $n({\bm r},t)$ and ${\bm v}({\bm r},t)$, which is present in Eq.~(\ref{eq:continuity}), and b) the non-linear term $[{\bm v}({\bm r}, t)\cdot \nabla]{\bm v}({\bm r}, t)$ in Eq.~(\ref{eq:NavierStokes}), representing the convective acceleration~\cite{landau06}.

Momentum-non-conserving collisions, such as those due to the friction of the electron liquid against the disorder potential, can be taken into account phenomenologically by adding a term of the type $-m_{\rm c} \gamma n({\bm r}, t){\bm v}({\bm r}, t)$ on the right-hand side of Eq.~(\ref{eq:NavierStokes}), where $\gamma$ is a damping rate~\cite{selberherr_book}. Furthermore, corrections to Eq.~(\ref{eq:NavierStokes}), stemming from the pseudo-relativistic nature of MDF flow in graphene, can be easily incorporated into the theory~\cite{tomadin_prb_2013,svintsov_prb_2013} and have been demonstrated to yield stronger rectified signals~\cite{tomadin_prb_2013}.

Finally, to close the set of equations, we need a relation between $\Phi({\bm r},t)$ and $n({\bm r},t)$. This depends on the screening exerted by dielectrics and conductors near the GPW.
If a metal gate is positioned underneath the GPW at a distance $d \ll W, k^{-1}$, where $k$ is the plasmon wave vector, the following local relation exists~\cite{tomadin_prb_2013}:
\begin{equation}\label{eq:LCA}
\Phi({\bm r}, t)\approx -\frac{e}{C}\delta n({\bm r},t)~,
\end{equation}
where $C = \epsilon_2/(4\pi d)$ is a capacitance per unit area and $\delta n({\bm r},t) \equiv n({\bm r}, t) -{\bar n}$.
Eq.~(\ref{eq:LCA}) greatly simplifies the theoretical analysis and, in fact, allows us to solve the problem in a fully analytical fashion~\cite{footnote:long-range-potential}, as we now detail.

Eqs.~(\ref{eq:continuity})-(\ref{eq:LCA}) need to be accompanied by boundary conditions. As explained in Appendix~A and in Ref.~\onlinecite{tomadin_arxiv_2014}, we fix $v_y(x,y=0,W)=0$ and $\partial_x v_y(x, y = 0,W)+\partial_y v_x(x, y = 0,W)=0$.

{\it Linear response theory and plasmons.}---The GPW supports collective charge density oscillations, i.e.~plasmons~\cite{Giuliani_and_Vignale}, which propagate along the ${\hat {\bm x}}$ direction and are confined in the ${\hat {\bm y}}$ direction. To calculate the frequency spectrum and potential profiles of these modes we have to linearize Eqs.~(\ref{eq:continuity})-(\ref{eq:NavierStokes}) and~(\ref{eq:LCA}). We write $n({\bm r},t) = {\bar n} + n_1({\bm r}, t) + n_2({\bm r}, t) + \dots$, ${\bm v}({\bm r},t) = {\bm v}_1({\bm r}, t) + {\bm v}_2({\bm r}, t) + \dots$, and $\Phi({\bm r},t) = \Phi_1({\bm r}, t) + \Phi_2({\bm r}, t) \dots$. Here $n_1({\bm r}, t)$, ${\bm v}_1({\bm r}, t)$, and $\Phi_1({\bm r}, t)$ [$n_2({\bm r}, t)$, ${\bm v}_2({\bm r}, t)$, and $\Phi_2({\bm r}, t)$] denote first-order [second-order] corrections with respect to equilibrium values (by ``equilibrium'' we here mean the state of the GPW in which a plasmon is not propagating). In the linearized theory we retain only terms of the first order. All the relevant details are reported in Appendix~B and C.

For the sake of simplicity, we assume a uniform equilibrium electron density in the GPW, disregarding the well known inhomogeneous doping ${\bar n} \to {\bar n}(y)$ that arises due to a back gate. Plasmons in back-gated waveguides, however, have been demonstrated~\cite{inhomogeneousdoping} to be similar to those of uniformly doped waveguides, provided that the Fermi energy is appropriately scaled to compensate for the singular behavior of the carrier density ${\bar n}(y)$ as $y \to 0, W$.

\begin{figure}
\includegraphics[width=\linewidth]{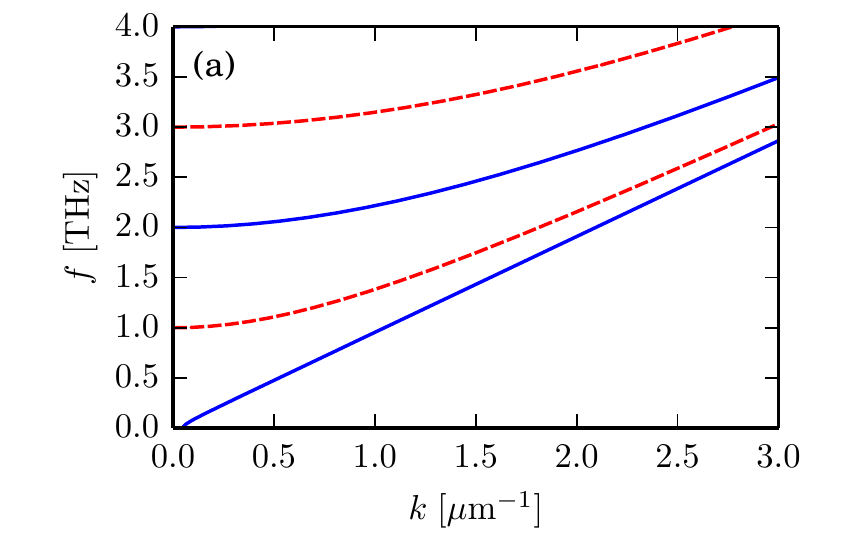}
\includegraphics[width=\linewidth]{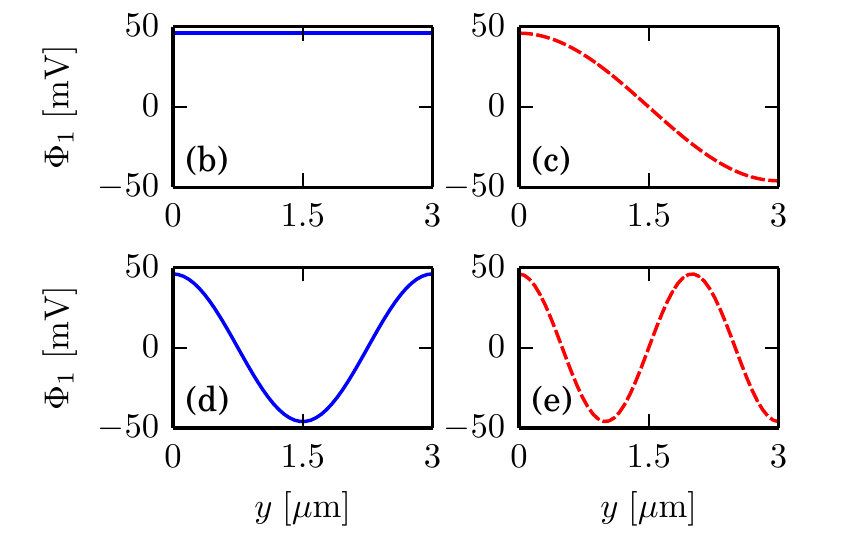}
\caption{(Color online) Panel (a): Dispersion relation $f_n(k) \equiv \omega_n(k)/(2\pi)$ of four low-energy plasmon modes ($n =0, \dots,3$) in a GPW with the following parameters: $W = 3~{\rm \mu m}$, $d = 100~{\rm nm}$, $\epsilon_2 = 3.9$, ${\bar n} = 10^{12}~{\rm cm}^{-2}$, $\gamma =0.3~ \Omega_0$, and $\nu=250~{\rm cm}^2/{\rm s}$. Solid (dashed) lines correspond to even (odd) modes. Panels (b)-(e): Corresponding electrical potential profiles $\Phi_1({\bm r}, t)$ evaluated at $x=0$ and $t = 0$ and plotted as functions of the transverse $y \in [0,W]$ coordinate. These results have been obtained by imposing electron density fluctuations equal to~\cite{footnote:amplitudeTHz} $\delta n/\bar{n}=1\%$.\label{fig:dispersion}}
\end{figure}

Plasmon modes are labelled by a wave number $k$ (stemming from translational invariance along the $\hat{\bm x}$ direction) and a discrete index $n = 0,1,2, \dots$. The associated ac electrical potential is given by
\begin{equation}\label{eq:ModesProfiles}
\Phi_1({\bm r}, t)=\varphi_n(y)e^{ikx-i\omega_n(k)t},
\end{equation}
where
\begin{equation}\label{eq:PhiFunctions}
\varphi_n(y)= \frac{1}{\sqrt{W}}\times\left\{
\begin{array}{l}
1,~{\rm for}~n=0\vspace{0.1cm}\\ \sqrt{2}\cos[n\pi y/W],~{\rm for}~n \neq 0
\end{array}
\right.~.
\end{equation}
The mode dispersion reads as following
\begin{equation}\label{eq:ViscousPlasmaWavesDispersion}
\omega_n(k) =  \sqrt{s^2 K^2_{n}- \frac{(\gamma + \nu K^2_{n})^2}{4}} - i \frac{\gamma + \nu K^2_{n}}{2}~,
\end{equation}
where $K^2_{n} = k^2 + q_n^2$, $q_{n} \equiv \pi n/W$, and $s= \sqrt{e^2 \bar{n}/(Cm_{\rm c})}$ is the hydrodynamic speed of sound.
It is useful to introduce the following natural frequency scale: $\Omega_0 \equiv s/W$. 
Setting external ($\gamma$) and internal ($\nu$) dissipation to zero in Eq.~(\ref{eq:ViscousPlasmaWavesDispersion}) we find the expected result $\omega_n(k) = s K_{n}$.
The lowest-energy $n=0$ mode shows an acoustic dispersion due to screening by the back gate.
Modes with $n \neq 0$ are gapped, i.e.~$\omega_n(k\to 0) = n \pi \Omega_0$.
The fundamental frequency is $\omega_{n=1}/(2\pi) = \Omega_0/2 \simeq 1.0~{\rm THz}$ for $W = 3~{\rm \mu m}$, $d = 100~{\rm nm}$, $\epsilon_2 = 3.9$, and ${\bar n} = 10^{12}~{\rm cm}^{-2}$. 
Dispersion relations and mode profiles for the above set of parameters are shown in Fig.~\ref{fig:dispersion}. 
In the approximation (\ref{eq:LCA}) the results do not depend on $\epsilon_{1}$.

When the $n$-th eigenmode of the GPW is excited by an external perturbation with frequency $\omega$, 
it propagates with a complex wave number
\begin{equation}\label{eq:wavevector}
k_n(\omega) = \sqrt{\frac{\omega^2+i \omega \gamma}{s^2 - i \nu \omega} - q^2_{n}}~.
\end{equation}
The wave number, $\Re e(k_n)$, and inverse damping ratio, $\gamma^{-1}_{n} \equiv \Re e(k_n)/\Im m(k_n)$, of the launched plasmon depend only on the excitation frequency $\omega$ and not on details of the tip-sample coupling~\cite{fei_nature_2012,chen_nature_2012,woessner_arxiv_2014,fei_nanolett_2011,alonso_science_2014}.
Physically, the dimensionless number $\gamma^{-1}_{n}$ controls the plasmon extinction length $\ell_{n} \equiv 1/\Im m(k_n) = \gamma^{-1}_{n}\lambda_{n}/(2\pi)$, with $\lambda_{n} = 2\pi/\Re e(k_{n})$ the plasmon wavelength.
With the value of $\nu$ given above and $\gamma = 0.3~\Omega_0$, the inverse damping ratio of the $n=0$ mode is $\gamma^{-1}_{0} \simeq 23$, while $\gamma^{-1}_{1} \simeq 7$ for $n=1$.

{\it Second-order theory of all-electrical detection.}---The rectified signal can be calculated by keeping track of the second-order terms $n_2({\bm r},t)$, ${\bm v}_2({\bm r},t)$, and $\Phi_2({\bm r},t)$ in the expansion of the hydrodynamic variables. Physically, the second-order response describes interactions between propagating modes. If only one mode propagates, a dc signal due to {\it self-mixing} of the plasmon field is generated. If more than one mode propagates, also interference terms will be generated. On general grounds~\cite{dyakonov_shur_theory}, we expect that the second-order response is composed by an oscillating component at frequency $2\omega$ (i.e.~second-harmonic generation) and a steady component. Since we are interested in detecting a dc signal, we can extract the rectified voltage $\delta V({\bm r})$ from the time average over one period of the external radiation of the second-order potential fluctuations: $\delta V({\bm r}) \equiv \langle\Phi_2({\bm r}, t)\rangle$. Averaging over time the second-order equations as explained in Appendix~B, we obtain
\begin{equation}\label{eq:averaged_cont_equation}
\nabla \cdot \delta {\bm v}({\bm r})=-\frac{1}{\bar{n}} \nabla \cdot \langle n_1({\bm r}, t){\bm v}_1({\bm r}, t)\rangle
\end{equation}
and
\begin{eqnarray}\label{eq:averaged_NS_equation}
&&-\frac{e}{m_{\rm c}}\nabla \delta V({\bm r})
+\gamma \delta {\bm v}({\bm r})-\nu \nabla^2\delta  {\bm v}({\bm r})\nonumber\\
&&=- \epsilon\langle [{\bm v}_1({\bm r}, t)\cdot\nabla] {\bm v}_1({\bm r}, t)\rangle-\frac{\nu}{\bar{n}}\langle n_1({\bm r}, t) \nabla^2  {\bm v}_1({\bm r}, t) \rangle~,\nonumber\\
\end{eqnarray}
where $\delta {\bm v} \equiv \langle{\bm v}_2({\bm r},t)\rangle$ is the time average of the velocity fluctuations. We urge the reader to note that in Eq.~(\ref{eq:averaged_NS_equation}) we have introduced a dimensionless parameter, $\epsilon$, which allows us to keep track of the role of different non-linearities in determining the rectified signal. By setting $\epsilon=0$ one neglects the convective non-linearity in the Navier-Stokes equation. Moreover, by setting $\epsilon=0$ {\it and} $\nu =0$, the Navier-Stokes equation reduces to the linearized Euler equation~\cite{landau06}, which leads to the {\it standard} Drude formula for the local conductivity. However, a finite rectified signal $\delta V({\bm r})$ exists in this case too and is entirely due to the non-linear $n({\bm r},t) {\bm v}({\bm r},t)$ coupling in the continuity equation.

Eqs.~(\ref{eq:averaged_cont_equation})-(\ref{eq:averaged_NS_equation}) are crucial since they relate the second-order quantities $\delta V$ and $\delta {\bm v}$ to the quantities $n_1({\bm r},t)$ and ${\bm v}_1({\bm r},t)$, which have been calculated in the linearized theory.
Furthermore, they can be used to calculate the dc signal $\delta V({\bm r})$ in response to plasmon propagation in any desired geometry.
As stated above, $\delta V({\bm r})$ can be measured by employing a set of ohmic contacts as in Fig.~\ref{fig:setup}.

We now evaluate $\delta V({\bm r})$ for the experimentally relevant case in which plasmons are launched at a specific location ${\bm r}^\star =(0,y^\star)$ with $y^\star \in [0,W]$ in the GPW.
The quantity $\delta V({\bm r})$ for $x \gg W$ and arbitrary $y$ can then be calculated according to the following procedure, which is typical of a scattering problem. a) For $x\gg W$ the plasmon velocity field can be written as a sum over propagating modes (i.e~modes with $\omega_n(k)<\omega$, where $\omega$ is the angular frequency of the stimulus that launches plasmons). All the other modes, which can be excited near ${\bm r}^\star$, exponentially damp out at large distances since they have a purely imaginary $k$---see Eq.~(\ref{eq:wavevector}). Furthermore, as shown in Appendix~C, the plasmon velocity field ${\bm v}_1$ is {\it irrotational} at large distances, i.e.~$\nabla \times {\bm v}_1 =0$ for $x \gg W$. Since $\nabla  \times \nabla \phi({\bm r}) \equiv 0$ for an arbitary scalar function $\phi({\bm r})$, we conclude that ${\bm v}_1({\bm r},t)$ for $x\gg W$ can be written as the gradient of a scalar function. In the language of scattering theory, we have built the so-called {\it asymptotic solution}, which we denote by  ${\bm v}^{({\rm a})}_1({\bm r},t)$. b) Let us imagine that an external perturbation with frequency $\omega$ launches, for example, an arbitrary linear combination with complex coefficients of the $n=0$ and $n=1$ GPW modes.
Because of a), we can write the corresponding asymptotic velocity field as
\begin{eqnarray}\label{eq:velocityfield}
{\bm v}^{({\rm a})}_1({\bm r}, t) &=& \frac{A}{2}\nabla[(1- \xi) \varphi_0(y)  e^{i {\bar k}_0 x}e^{- \beta_0 x} \nonumber \\
&+&\xi e^{i \alpha} \varphi_1(y)  e^{i {\bar k}_1x}e^{- \beta_1 x}] e^{-i\omega t} +{\rm c.c.}~,
\end{eqnarray}
where $x>0$ and the functions $\varphi_n(y)$  have been introduced earlier in Eq.~(\ref{eq:PhiFunctions}).
In Eq.~(\ref{eq:velocityfield}) $A = {\bar v} W^{3/2}$ is an unknown amplitude (here ${\bar v}$ has physical dimensions of a velocity), which can be estimated as discussed below, ${\bar k}_n = \Re e[k_n(\omega)]$ and $\beta_n = \Im m[k_n(\omega)]$ with $k_n(\omega)$ as in Eq.~(\ref{eq:wavevector}), $\xi \in [0,1]$ is a real parameter that allows us to interpolate between the case in which only the $n=0$ mode is launched ($\xi = 0$) and the case in which only the $n=1$ mode is launched ($\xi = 1$), and $e^{i\alpha}$ (with $\alpha$ real) is the relative phase between the two modes. In the case in which plasmons are launched at ${\bm r}^\star$ by using a metallized tip illuminated by light~\cite{fei_nature_2012,chen_nature_2012,woessner_arxiv_2014,fei_nanolett_2011,alonso_science_2014}, $\xi$ depends on the tip-sample coupling: for example, for a tip placed symmetrically with respect to the GPW axis $\xi$ vanishes. 
In practice, $\xi$ can be varied by moving the tip along the $\hat{\bm y}$ direction.
c) With the velocity field in Eq.~(\ref{eq:velocityfield}), one can easily calculate the asymptotic density profile $n^{({\rm a})}_1({\bm r},t)$ from the continuity equation. 
d) The quantities $n^{({\rm a})}_1({\bm r},t)$ and ${\bm v}^{({\rm a})}_1({\bm r}, t)$ are then used to calculate the temporal averages that appear on the right-hand side of Eqs.~(\ref{eq:averaged_cont_equation})-(\ref{eq:averaged_NS_equation}). 
e) Finally, $\delta V({\bm r})$ is found by solving Eqs.~(\ref{eq:averaged_cont_equation})-(\ref{eq:averaged_NS_equation}).

\begin{figure}
\begin{overpic}[width=\linewidth]{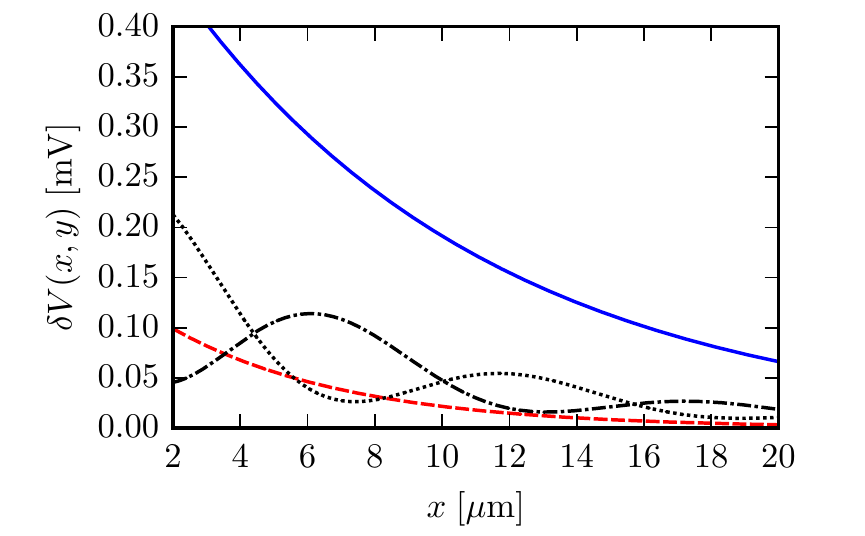}\put(2,60){(a)}\end{overpic}
\begin{overpic}[width=\linewidth]{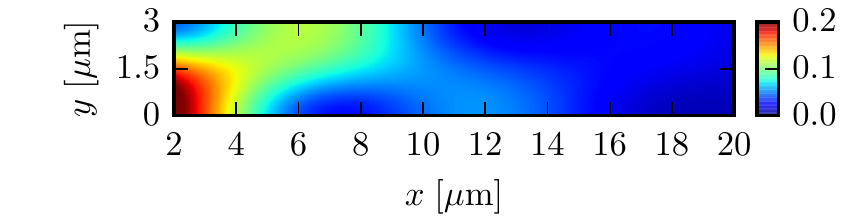}\put(2,25){(b)}\end{overpic}
\caption{\label{fig:dcsignal}
(Color online)
Panel (a) The dc potential $\delta V({\bm r})$ (in millivolts) as a function of $x$ (in $\mu{\rm m}$), calculated at the edges of the GPW, i.e.~at $y = 0$ (solid, dashed, and dotted lines) and $y = W$ (dash-dotted line).
These numerical results have been calculated by setting the following parameter values in Eq.~(\ref{eq:velocityfield}): $\xi = 0$ and $\alpha = 0$ (solid line), $\xi = 1$ and $\alpha = 0$ (dashed line), $\xi = 1/2$ and $\alpha = \pi/4$ (dotted and dash-dotted lines).
The other parameters are as in Fig.~\ref{fig:dispersion}.
Finally, we have taken $\omega / (2 \pi) \simeq 1.2~{\rm THz}$ in Eq.~(\ref{eq:wavevector}).
Note that $\omega/(2\pi)$ is $20\%$ larger than the fundamental frequency for the laser to be able to excite the two lowest modes of the GPW.
The scale is in millivolts.
In all cases the value of $A$ has been chosen to yield $\delta n/{\bar n} = 1\%$.
This normalization implies different values of $A$ for different values of $\xi,\alpha$.
Panel (b) Spatial map of the dc potential $\delta V({\bm r})$ calculated by setting $\xi = 1/2$ and $\alpha = \pi/4$.
The color bar shows the potential amplitude in millivolts. All other parameters are as in panel (a).}
\end{figure}

Simple and compact analytical expressions, obtained by following steps a)-e), are available for $\delta V({\bm r})$ in the extreme cases $\xi=0$ and $\xi=1$ and are presented Appendix~D. In the general 
$\xi \neq 0,1$ case an oscillatory term with spatial periodicity $2\pi/({\bar k}_0 - {\bar k}_1)$ appears along the ${\hat {\bm x}}$ direction due to interference of the two modes in Eq.~(\ref{eq:velocityfield}).
Illustrative numerical results can be found in Fig.~\ref{fig:dcsignal}.
Notice that the dc signal is $\lesssim 0.5~{\rm mV}$ and that its spatial extension is $\sim 20~\mu{\rm m}$.
The dc potentials on the top and bottom GPW edges are not equal in the case $\xi \neq 0,1$ since Eq.~(\ref{eq:velocityfield}) is a superposition of modes with different parity.
The quantity $A$ in Eq.~(\ref{eq:velocityfield}) was estimated with reference to Refs.~\onlinecite{fei_nature_2012,chen_nature_2012,woessner_arxiv_2014,fei_nanolett_2011,alonso_science_2014}, where a metallized tip is used to launch plasmons.
In this case, $A$ can be calculated starting from the amplitude of density oscillations $\delta n$ in units of ${\bar n}$, created by the tip at ${\bm r}^\star$.
The results in Fig.~\ref{fig:dcsignal} have been produced by using~\cite{footnote:amplitudeTHz} $\delta n/{\bar n} = 1\%$ at $x=0$. This ``normalization" condition yields {\it different} values of $A$ for different values of $\xi,\alpha$.
In other words, $A$ measures how well the tip couples to the linear combination of modes in Eq.~(\ref{eq:velocityfield}).
The $\xi =0$ mode has a better coupling to the tip (and therefore yields a larger dc signal) than the linear superposition of $n=0$ and $n = 1$ modes corresponding to $\xi =1/2$ and $\alpha=\pi/4$.

{\it Conclusions.}---In summary, we have discussed an architecture based on a graphene waveguide where electrical detection of plasmons may be experimentally achieved.
We have theoretically demonstrated that rectification of the ac field of a propagating plasmon, which is enabled by non-linear terms in Eqs.~(\ref{eq:continuity})-(\ref{eq:LCA}), yields a spatially-dependent dc signal $\delta V({\bm r})$.
The experimental exploitation of similar non-linearities has recently led~\cite{THzdetectors} to room-temperature graphene THz photodetectors.
We stress that $\delta V({\bm r})$ can be calculated from Eqs.~(\ref{eq:averaged_cont_equation})-(\ref{eq:averaged_NS_equation}) and can be measured by using lateral probe contacts as in Fig.~\ref{fig:setup}.
Simple analytical expressions for $\delta V({\bm r})$ have been given in Appendix~D for the cases $\xi =0,1$.
Numerical results for the general case $\xi \neq 0,1$ have been presented in Fig.~\ref{fig:dcsignal}.
Such values of dc voltages can be easily measured.

\acknowledgments
We wish to thank R.~Hillenbrand, F.~Koppens, A.~Principi, D.~Spirito,  A.~Tredicucci, and especially M.~Lundeberg for many useful discussions.
This work was supported by the EC under the Graphene Flagship program (contract no.~CNECT-ICT-604391) and MIUR through the programs ``FIRB - Futuro in Ricerca 2010'' - Project ``PLASMOGRAPH'' (Grant No.~RBFR10M5BT) and ``Progetti Premiali 2012" - Project ``ABNANOTECH''.

\appendix

\section{Boundary conditions}

Hydrodynamic equations need to be accompanied by appropriate boundary conditions (BCs) at the boundaries of the region occupied by the electron fluid. In the general case of viscous flow, we need BCs on the normal and tangential components of the current evaluated at the boundaries of the GPW.

The BC on the normal component can be simply derived by requiring that no charge exit the boundary of the system. The normal component of the current ${\bm J} = n {\bm v}$ evaluated at the boundaries must therefore vanish
\begin{equation}\label{eq:normalBC}
\hat{n}_{i} J_{i}=n \hat{n}_{i} v_{i}=0~,
\end{equation}
where $\hat{\bm n}$ is a unit vector normal to the boundary. In our simple rectangular geometry this translates to
\begin{equation}\label{eq:normalBC2}
\left.v_y(x,y)\right|_{y=0,W} =0~.
\end{equation}

In the description of the motion of liquids near a fixed surface, the following BC, which physically corresponds to the vanishing of the tangential component of the velocity field, is usually assumed~\cite{landau06}: 
\begin{equation}\label{eq:tangentialBC}
\epsilon_{ij} \hat{n}_i v_j=0~.
\end{equation}
where $\epsilon_{ij}$ is the 2D fully-antisymmetric tensor. Eq.~(\ref{eq:tangentialBC}) can be justified for molecular liquids~\cite{landau06}, where interactions between two molecules of the liquid are essentially of the same type of the interaction between a molecule in the liquid and one in the surface of the container.

For the electron liquid in graphene, however, in which boundaries are defined by etching or electrostatic gating, we cannot identify a mechanism through which the boundary can exert a tangential force on the liquid.
For this reason we chose to apply the following BC~\cite{tomadin_arxiv_2014}:
\begin{equation}\label{eq:tangentialBC2}
\epsilon_{ij} \hat{n}_i \sigma_{jk}' \hat{n}_k=0~.
\end{equation}
Physically, the previous equation represents the vanishing of the tangential force on the liquid, and is commonly used~\cite{landau06} for liquids near a free surface.
In Eq.~(\ref{eq:tangentialBC2}) $\sigma_{jk}'$ is the 2D viscous stress tensor, i.e.~$\sigma_{jk}'=\eta \left(\partial_j v_k+ \partial_k v_j-\delta_{jk}\partial_l v_l\right)$. In writing the previous expression for $\sigma_{jk}'$ we have set to zero a contribution due to the bulk viscosity $\zeta$.

In our geometry, Eq.~(\ref{eq:tangentialBC2}) reduces to:
\begin{equation}\label{eq:tangentialBC3}
\left[\partial_x v_y(x, y)+\partial_y v_x(x, y)\right]_{y=0,W}=0~,
\end{equation}
as reported in the main text.

\section{Perturbative solution of the hydrodynamic equations}

In the main text we have solved to the coupled non-linear equations (1)-(2) by employing the following perturbative ansatz~\cite{dyakonov_shur_theory}:
\begin{equation}\label{eq:PerturbativeAnsatz}
\begin{split}
&n({\bm r}, t)= \bar{n} + \lambda n_1({\bm r}, t) + \lambda^2 n_2({\bm r}, t) + \dots\\
& {\bm v}({\bm r}, t) = \lambda {\bm v}_1({\bm r}, t) + \lambda^2 {\bm v}_2({\bm r}, t) + \dots\\
&\Phi({\bm r}, t ) = \lambda \Phi_1({\bm r}, t) + \lambda^2 \Phi_2({\bm r}, t) + \dots\\
\end{split}~,
\end{equation}
where $\lambda$ is a bookkeeping parameter, which will be set to unity at the end of calculations.

Inserting the above expressions into Eqs.~(1)-(2) of the main text and retaining only terms of the first-order in $\lambda$, 
we obtain the linearized continuity equation,
\begin{equation}\label{eq:continuitylinearized}
\partial_t n_1({\bm r}, t) + \nabla \cdot [\bar{n}{\bm v}_1({\bm r}, t)]=0~,
\end{equation}
and the linearized Navier-Stokes equation,
\begin{equation}\label{eq:linearizedNavierStokes}
\partial_t {\bm v}_1({\bm r}, t) = \frac{e}{m_{\rm c}} \nabla \Phi_1({\bm r}, t) - \gamma {\bm v}_1({\bm r}, t) + \nu \nabla^2 {\bm v}_1({\bm r}, t)~.
\end{equation}

\begin{widetext}

These two equations have been used in the main text to calculate the spectrum of plasmons in the GPW. Note that, if the last equation is written in Fourier transform with respect to both space and time, it leads to a conductivity of the form
\begin{equation}\label{eq:conductivity}
\sigma(q,\omega)  = \frac{i {\cal D}_0/\pi}{\omega + i\gamma + i \nu q^2}~, 
\end{equation}
where ${\cal D}_0 = \pi {\bar n} e^2/m_{\rm c}$ is the (non-interacting) Drude weight~\cite{galileaninvariance}. 
Viscosity yields a non-local correction to the ordinary Drude formula.

At second order in $\lambda$ we obtain the following equations:
\begin{equation}\label{eq:continuitysecondorder}
\partial_t n_2({\bm r}, t)+\nabla \cdot \left[ \bar{n}{\bm v}_2({\bm r}, t)+n_1({\bm r}, t){\bm v}_1({\bm r}, t)\right]=0
\end{equation}
and
\begin{eqnarray}\label{eq:NavierStokessecondorder}
&&m_{\rm c} \bar{n}\left[\partial_t {\bm v}_2({\bm r};t)+({\bm v}_1({\bm r};t)\cdot\nabla) {\bm v}_1({\bm r};t)\right]
+m_{\rm c}n_1({\bm r};t)\partial_t {\bm v}_1({\bm r};t)
=e \bar{n}\nabla \Phi_2({\bm r};t)+e n_1({\bm r};t) \nabla \Phi_1({\bm r};t)\nonumber \\
&&-m_{\rm c} \gamma \bar{n}{\bm v}_2({\bm r};t)-m_{\rm c} \gamma n_1({\bm r};t) {\bm v}_1({\bm r};t)+\eta \nabla^2 {\bm v}_2({\bm r};t)~.\nonumber\\
\end{eqnarray}
The latter can be simplified thanks to the linearized Navier-Stokes equation (\ref{eq:linearizedNavierStokes}), leading to:
\begin{equation}\label{eq:NavierStokessecondorder2}
\partial_t {\bm v}_2({\bm r}, t)-\frac{e}{m_{\rm c}}\nabla \Phi_2({\bm r}, t)+ \gamma {\bm v}_2({\bm r}, t)-\nu \nabla^2 {\bm v}_2({\bm r}, t)
=-\nu \frac{n_1({\bm r}, t)}{\bar{n}}\nabla^2  {\bm v}_1({\bm r}, t) - \epsilon [{\bm v}_1({\bm r}, t)\cdot \nabla] {\bm v}_1({\bm r}, t)~.
\end{equation}
Averaging over time both sides of Eqs.~(\ref{eq:continuitysecondorder}) and~(\ref{eq:NavierStokessecondorder2}) we finally find Eqs.~(9)-(10) of the main text.

\section{Normal modes of oscillation: plasmons}

In this Section we use the linearized hydrodynamic equations (\ref{eq:continuitylinearized})-(\ref{eq:linearizedNavierStokes}) to show that the GPW supports collective charge density oscillations, i.e.~plasmons.

We look for oscillating solutions where all the quantities vary in time as $f({\bm r},t) = {\rm Re}[ e^{-i \omega t}f_{\omega}({\bm r})]$, 
where $\omega$ is, in general, a complex number (to allow for modes with a finite linewidth).
Eqs.~(\ref{eq:continuitylinearized})-(\ref{eq:linearizedNavierStokes}) then transform into:
\begin{equation}\label{eq:ContinuityOmega}
-i \omega n_{1,\omega}({\bm r}) + \nabla \cdot [\bar{n}{\bm v}_{1,\omega}({\bm r})]=0
\end{equation}
and
\begin{equation}\label{eq:NavierStokesOmega}
-i \omega {\bm v}_{1,\omega}({\bm r}) = -\frac{e^2}{m_{\rm c}C} \nabla n_{1,\omega}({\bm r}) - \gamma {\bm v}_{1,\omega}({\bm r}) + \nu \nabla^2 {\bm v}_{1,\omega}({\bm r})~,
\end{equation}
where the electric potential $\Phi_1$ has been rewritten by emplying the local capacitance approximation, i.e.~Eq.~(4) of the main text.

We can eliminate $n_{1,\omega}({\bm r})$ in Eq.~(\ref{eq:NavierStokesOmega}) by using Eq.~(\ref{eq:ContinuityOmega}). We find
\begin{equation}\label{eq:ModesEquation}
\omega^2 {\bm v}_{1,\omega}({\bm r}) + \frac{e^2}{m_{\rm c}C} \nabla \left[ \nabla \cdot {\bm v}_{1,\omega}({\bm r})\right] + i \omega \gamma {\bm v}_{1,\omega}({\bm r}) - i \omega \nu \nabla^2 {\bm v}_{1,\omega}({\bm r})~=0.
\end{equation}
We can take advantage of translational invariance along the $\hat{\bm x}$ direction by introducing the following ansatz for ${\bm v}_{1,\omega}({\bm r})$:
\begin{equation}\label{eq:VelocityAnsatz}
{\bm v}_{1,\omega}({\bm r})=e^{ikx}\left[\hat{\bm x}\sum_{n=0}^{\infty} a_n \varphi_n(y)+\hat{\bm y} \sum_{n=1}^{\infty} b_n \psi_n(y) \right]~,
\end{equation}
where the functions $\varphi_n$ have been introduced in the main text, while 
$\psi_n(y)=\sqrt{2/W} \sin (n \pi y/W)$. It can be easily checked that $\{\varphi_n(y), n =0,1,\dots\}$ and $\{\psi_n(y), n = 1,\dots\}$ represent 
complete sets of orthonormal functions in the interval $y\in [0,W]$. Furthermore, the ansatz (\ref{eq:VelocityAnsatz}) respects the BCs~(\ref{eq:normalBC2}) and~(\ref{eq:tangentialBC3}).

Substituting Eq.~(\ref{eq:VelocityAnsatz}) in Eq.~(\ref{eq:ModesEquation}) and using standard completeness theorems, we can transform Eq.~(\ref{eq:ModesEquation}) into a set of decoupled matrix equations, one for each $n$:
\begin{equation}\label{eq:Matrixequation}
\left[\omega^2+i\gamma \omega + i \omega \nu(k^2+n^2\pi^2/W^2)\right]
\left(
\begin{array}{c}
a_n\\
b_n\\
\end{array}
\right)
=\Omega^2_{0}
\left(
\begin{array}{cc}
k^2W^2 & -ikWn\pi\\
ikWn\pi & n^2\pi^2\\
\end{array}
\right)
\left(
\begin{array}{c}
a_n\\
b_n\\
\end{array}\right)~,
\end{equation}
where $\Omega_0^2 \equiv e^2\bar{n}/(Cm_{\rm c} W^2)$ is the square of a natural frequency scale, as in the main text.

The $2\times 2$ matrix on the right-hand side of Eq.~(\ref{eq:Matrixequation}) has the following eigenvectors
\begin{equation}\label{eq:Eigenvectors}
{\bm \Phi}_n\equiv \left(
\begin{array}{c}
ik\\
-n \pi /W\\
\end{array}
\right)
\;;
{\bm \Psi}_n\equiv \left(
\begin{array}{c}
i n \pi/W\\
k\\
\end{array}
\right)~.
\end{equation}
The solutions are then given by Eq.~(\ref{eq:VelocityAnsatz}) where, for every $n' =0,1,\dots$, the vector 
$(a_{n},b_{n}) \equiv \delta_{n,n'} {\bm \Phi}^{\rm t}_n$ or $(a_{n},b_{n}) \equiv \delta_{n,n'} {\bm \Psi}^{\rm t}_n$.

The solutions built by using ${\bm \Phi}_n$ in Eq.~(\ref{eq:Eigenvectors}), which has eigenvalue $k^2W^2+n^2\pi^2$, lead to an {\it irrotational} spatial profile of the type:
\begin{equation}\label{eq:VelocityField}
{\bm v}_{1,\omega}({\bm r})=Ae^{ikx}\left[ik \varphi_n(y) \hat{\bm x} - n\pi \psi_n(y) \hat{\bm y}\right]=A\nabla\left[e^{ikx}\varphi_n(y)\right]~,
\end{equation}
with $\omega = \omega_{n}(k)$ given by the dispersion relation in Eq.~(7) of the main text.

The second eigenvector, ${\bm \Psi}_n$, with zero eigenvalue, leads to {\it solenoidal} solutions with no density fluctuations.
These modes have a purely imaginary frequency (because they do not experience a restoring force that sustains them) and cannot therefore propagate along the GPW. Even if they are excited by an external perturbation (note that an external scalar potential does not couple to these modes) they decay exponentially for $x\gg W$.
We can therefore safely exclude them from the construction of the asymptotic solution ${\bm v}^{({\rm a})}_1({\bm r},t)$ in the main text.

\section{Analytical results for the rectified signal}

We here report our main analytical results for the rectified signal in the extreme cases $\xi =0$ and $\xi =1$---see Eq.~(11) in the main text. For $\xi =0$ we find
\begin{equation}\label{eq:solutionsn0}
\delta V({\bm r})= {\cal V} e^{-2\beta_0 x} W^2\frac{({\bar k}^2_0 + \beta^2_0) \left[\epsilon \omega \beta_0 + {\bar k}_0 \gamma + \nu {\bar k}_0 \left({\bar k}^2_0 - 3 \beta^2_0 \right)\right]}{4\beta_0\omega}
\end{equation}
while, for $\xi =1$,
\begin{eqnarray}\label{eq:solutionsn1}
\delta V({\bm r}) &=&{\cal V} e^{-2\beta_1 x}W^2\Bigg\{\frac{\left(\epsilon\omega \beta^2_1+\beta_1 {\bar k}_1 \gamma\right)\left(\beta^2_1 +{\bar k}^2_1+ q^2_{1}\right) + \nu \beta_1 {\bar k}_1\left({\bar k}^4_1 - 3\beta^4_1 + q^4_{1} - 2{\bar k}^2_1 \beta^2_1 -6q^2_{1}\beta^2_1 + 2q^2_{1} {\bar k}^2_1\right)}{4\beta^2_1 \omega}\nonumber\\
&+&\frac{\varphi_2(y)}{\sqrt{2}}\sqrt{W}\frac{\left[\epsilon\omega(\beta^2_1 - q^2_{1})+\beta_1 {\bar k}_1 \gamma\right]\left(\beta^2_1  + {\bar k}^2_1 - q^2_{1}\right) + \nu \beta_1 {\bar k}_1\left({\bar k}^4_1 - 3\beta^4_1 - 3q^4_{1} - 2k^2_{1}\beta^2_{1} + 6 q^2_{1} \beta^2_{1} + 6 q^2_{1} k^2_{1}\right)}{4\omega(\beta^2_{1} - q^2_{1})}\Bigg\}~.\nonumber\\
\end{eqnarray}

\end{widetext}

Here ${\cal V} = m_{\rm c} |A|^2 W^{-3}/e = m_{\rm c} {\bar v}^2/e$ has physical dimensions of a voltage.
Once again, we hasten to stress that a finite value of $\delta V({\bm r})$ exists also in the ordinary Drude $\nu=\epsilon = 0$ limit, as it can be explicitly checked from Eqs.~(\ref{eq:solutionsn0})-(\ref{eq:solutionsn1}). In this case the rectified signal is due to the hydrodynamic $n({\bm r}, t){\bm v}({\bm r},t)$ non-linear coupling in the continuity equation.

\end{document}